\documentclass[oupdraft]{bio}
\usepackage{url}
\usepackage{siunitx}
\usepackage{pdflscape}
\usepackage{multirow}
\usepackage{placeins}
\usepackage{dsfont,bm}
\usepackage{rotating}
\usepackage[hmarginratio=1:1]{geometry}

\history{Received February 23, 2019}

\begin{document}

\title{Statistical methods for biomarker data pooled from multiple nested case-control studies}

\author{ABIGAIL SLOAN*\\
\textit{Department of Biostatistics, Harvard T. H. Chan School of Public Health, Boston, MA, 02115} \\ \vspace{3mm}
MOLIN WANG* \\ \vspace{2mm}
\textit{Departments of Biostatistics and Epidemiology, Harvard T. H. Chan School of Public Health, \\
Channing Division of Network Medicine, Department of Medicine, Brigham and Women's Hospital, and Harvard Medical School, Boston, MA, 02115}
\\[2pt]
\vspace{3mm}
{asloan@g.harvard.edu \& stmow@channing.harvard.edu}}

\markboth%
{Sloan and Wang}
{Methods for Combining Biomarker Data for Case-Control Studies}

\maketitle

\footnotetext{To whom correspondence should be addressed.}

\begin{abstract}
{Pooling biomarker data across multiple studies allows for examination of a wider exposure range than generally possible in individual studies, evaluation of population subgroups and disease subtypes with more statistical power, and more precise estimation of biomarker-disease associations. However, biomarker measurements often require calibration to a reference assay prior to pooling due to assay and laboratory variability across studies. We propose several methods for calibrating and combining biomarker data from nested case-control studies when reference assay data are obtained from a subset of controls in each contributing study. Specifically, we describe a two-stage method and two aggregated methods, named the internalized and full calibration methods, to evaluate the main effect of the biomarker exposure on disease risk and whether that association is modified by a potential covariate. The internalized method uses the reference laboratory measurement in the analysis when available and otherwise uses the calibrated measurement. The full calibration method uses calibrated biomarker measurements for all subjects, even those with reference laboratory measurements. Our results demonstrate that the full calibration method is the preferred aggregated approach to minimize bias in point estimates regardless of the inclusion of an interaction term in the model. We also observe that the two-stage and full calibration methods provide similar effect and variance estimates, but that the variance estimates for these two methods are slightly larger than those from the internalized approach. As an illustrative example, we apply the methods in a pooling project of nested case-control studies to evaluate (i) the association between circulating vitamin D levels and risk of stroke, and (ii) how BMI modifies the association between circulating vitamin D levels and cardiovascular disease.}
{Aggregation; Biomarker; Calibration; Conditional logistic regression; Pooling project; Nested case-control study}
\end{abstract}

\section{Introduction}

Combining data from multiple studies to maximize sample size has become a common strategy to quantify exposure-disease associations, including those where the exposure is a biomarker. The increased sample sizes facilitate subgroup analyses, allow more precise estimation of the biomarker exposure effect over a wider range of biomarker measurements, and avoid issues related to data sparsity \citep{key2010pooling,smith2006methods}. The increase in the use of pooling consortia over time reflects the advantages of big data in epidemiology and its promises to improve quantification of disease risk factors. Note that we use the term pooling throughout this paper to refer to data combination and not physical specimen combination.  
Here, we define biomarkers as measurable indicators of health at the molecular, biochemical, or cellular level \citep{key2010pooling}. Examples include proteins, antibodies, hormones, and blood cholesterol. Many existing pooling consortia analyze biomarker-disease associations, including the Cohort Consortium Vitamin D Pooling Project of Rarer Cancers \citep{gallicchio2010circulating}, the Endogenous Hormones, Nutritional Biomarkers, and Prostate Cancer Collaborative Group \citep{key2015carotenoids}, the COPD Biomarkers Qualification Consortium Database \citep{tabberer2017copd}, and the Circulating Biomarkers and Breast and Colorectal Cancer Consortium \citep{mccullough2018circulating}, among others. 

In order to pool biomarker data, investigators must address potential between-study variation in biomarker data due to assay or laboratory variability if not all samples are assayed at the same laboratory with the same assay near the same time. For example, estradiol, testosterone, and insulin-like growth factor 1 have highly variable measurements across assays \citep{key2010pooling,tworoger2006use}. Measurements of blood-circulating vitamin D (25(OH)D) also vary up to 40\% between laboratories and assays \citep{barake201225,lai2012variability,snellman2010determining}. Consequently, biomarker measurements from studies conducted using different laboratories and assays must be harmonized before pooled analyses occur if the goal is to examine absolute concentrations using the same metric across studies. In practice, we standardize biomarker measurements by defining study-specific calibration models. After re-assaying a random subset of noncase biospecimens from each study at a designated reference laboratory, we estimate study-specific models between the original ``local" laboratory measurements and the reference laboratory measurements. These calibration models are then used to estimate the reference laboratory biomarker measurement among the remaining subjects in each study. Following this calibration procedure, the standardized biomarker measurements can be used in statistical analyses. In practice, re-assayed biospecimens are typically selected at random from controls in each study owing to concerns about the availability of case biospecimens \citep{sloan2018design}.

Two pooling methods exist for analyzing data from multiple studies, namely the two-stage approach and aggregated approach \citep{smith2006methods, debray2013individual}. Under the two-stage method, investigators complete study-specific analyses in the first stage followed by meta-analysis in the second stage. In the second approach, investigators combine data from all participating cohorts before performing statistical analyses on the collective dataset. \citet{sloan2018design} focused on cohort studies and subdivided the aggregated approach into the internalized and full calibration approach. The internalized method uses the reference laboratory measurement in the analysis when available and uses the calibrated measurement otherwise. In contrast, the full calibration method uses calibrated biomarker measurements exclusively for all subjects regardless of the availability of reference laboratory measurements. In this paper, we derive these approaches under the paradigm of matched and nested case-control studies, allowing the potential inclusion of a  biomarker-covariate interaction term.

We can equivalently view pooled and calibrated biomarker data as a covariate measurement error problem. If we treat the reference and local laboratory measurements as the true and surrogate biomarker values respectively \citep{Carroll:1012043}, we can envision each study-specific calibration model as a different measurement error model. We leverage an existing strategy in the measurement error literature, namely regression calibration \citep{Carroll:1012043,rosner1990correction}, to form the basis of our methods. Although we categorize each method as a two-stage or aggregated approach, all methods utilize concepts underlying regression calibration.

In this paper, we propose calibration methods for pooled biomarker data from matched and nested case-control studies that allow inference on the main effect of the biomarker in addition to biomarker-covariate interaction terms. Section 2 presents the models and statistical methods. Section 3 compares the methods via simulation, stratifying by the inclusion of a covariate-biomarker interaction term. Section 4 illustrates the methods in several examples involving 25(OH)D data pooled from the Nurses' Health Study I (NHS1), Nurses' Health Study II (NHS2), and Health Professionals Follow-up Study (HPFS) for stroke and cardiovascular disease (CVD) outcomes. Section 5 discusses our results.

\section{Methods}

\subsection{Model and approximate conditional likelihood}

Let $s=1,\dots,S$ index the studies contributing to the pooled analysis, where the first $Q$ do not use the reference laboratory to measure the biomarker of interest. For presentational simplicity, we focus on $1:m_s$ matching; it is straightforward to extend the methods to settings where multiple cases exist in some or all strata. Allow subscript $i$ to index individuals within stratum $j$ of study $s$ such that $i=m_s$ corresponds to the case and $i=0,1,\dotsc,m_s-1$ correspond to the matched controls. 

Let $Y_{sji}$ be the binary disease outcome, $X_{sji}$ be the biomarker exposure measurement from the reference lab, $W_{sji}$ be the biomarker exposure measurement from the local lab, and $\bm{Z}_{sji}$ be a vector of other covariates. Without further specification, all vectors are column vectors. We assume $X_{sji}$ is not available for all subjects due to the usage of a local laboratory to obtain biomarker measurements in at least one study. However, we assume $W_{sji}$ is available for all subjects in studies obtaining measurements from local labs. Using $W_{sji}$ as a direct substitute for $X_{sji}$ in the analysis will lead to biased estimates of the biomarker-outcome relationship if measurement variability exists across the study laboratories.

To estimate the biomarker exposure effect under the aggregated approach, we develop a likelihood-based method. The conditional logistic regression model for the biomarker-disease association is 

\begin{equation} \label{logit}
\text{logit}(P(Y_{sji}=1|X_{sji}, \bm{Z}_{sji}))= \beta_{0sj} + \beta_xX_{sji}+\bm{\beta}_z^T \bm{Z}_{sji},
\end{equation}

\noindent where $\beta_{0sj}$ is a stratum-specific intercept and $\bm{\beta}_z$ is a vector of covariate effects. We aim to provide point and interval estimates for $\beta_x$. For nested case-control studies using incidence density sampling, $\beta_x$ is the log relative risk (RR) describing the biomarker-disease association \citep{prentice1978retrospective}. 

Let vectors $\bm{X}_{sj}$, $\bm{W}_{sj}$, and matrix $\bm{Z}_{sj}$ contain their respective measurements from all individuals in stratum $j$ of study $s$. The conditional likelihood \citep{breslow1978estimation} is

\begin{equation*}
\begin{split}
L & = \prod_s \prod_j P(Y_{sj0}=0, \dots ,Y_{sj(m_s-1)}=0, Y_{sjm_s}=1|\bm{X}_{sj}, \bm{Z}_{sj}, \sum_{i=0}^{m_s} Y_{sji}=1) \\ 
& = \prod_s \prod_j \frac{\exp\{\beta_{0sj}+\beta_x X_{sjm_s}+\bm{\beta}_z^T\bm{Z}_{sjm_s} \} }{\sum_{i=0}^{m_s} \exp\{\beta_{0sj}+\beta_x X_{sji}+ \bm{\beta}_z^T \bm{Z}_{sji} \}}\\
& = \prod_s \prod_j \left(1+ \sum_{i=0}^{m_s-1} \exp \{ \beta_x(X_{sji}-X_{sjm_s})  +\bm{\beta}_z^T(\bm{Z}_{sji} -\bm{Z}_{sjm_s} )\} \right)^{-1}. \\
\end{split}
\end{equation*}

\noindent The conditional likelihood contribution cannot be computed for studies using a local laboratory because $X_{sji}$ is unavailable for individuals outside the calibration subset. Under a surrogacy assumption, we derive an approximate conditional likelihood below that uses an estimate of $X_{sji}$ in place of $X_{sji}$ itself. This surrogacy assumption states that $f(Y_{sji}|X_{sji}, \bm{Z}_{si},W_{si},\sum_{i=1}^{m_s} Y_{sji}=1) = f(Y_{sji}|X_{sji},\bm{Z}_{sji},\sum_{i=1}^{m_s} Y_{sji}=1)$; in essence, the reference laboratory measurement provides at least as much information about the likelihood of disease as the local laboratory measurement, conditional on the covariates and matching scheme.

Under aggregation, the likelihood contribution from a stratum with only local laboratory biomarker measurements is

\begin{equation*}
\begin{split}
l_{sj} & = P(Y_{sj0}=0, \dots, Y_{sjm_s}=1 |  \bm{W}_{sj},\bm{Z}_{sj}, \sum_{i=0}^{m_s} Y_{sji} =1) \\
& = \int P (Y_{sj0}=0., \dots , Y_{sjm_s}=1| \bm{X}_{sj}, \bm{W}_{sj}, \bm{Z}_{sj},\sum_{i=0}^{m_s} Y_{sji} =1 ) f(\bm{X}_{sj}| \bm{W}_{sj}, \bm{Z}_{sj},\sum_{i=0}^{m_s} Y_{sji} =1) d \bm{X}_{sj} \\
& = \int \left(1+ \sum_{i=0}^{m_s-1} \exp \{ \beta_x(X_{sji}-X_{sjm_s}) + \bm{\beta}_z^T(\bm{Z}_{sji} -\bm{Z}_{sjm_s})\} \right)^{-1} f(\bm{X}_{sj}|\bm{W}_{sj},\bm{Z}_{sj},\sum_{i=0}^{m_s} Y_{sji} =1) d \bm{X}_{sj} \\
& = E_{\bm{X}_{sj}| \bm{W}_{sj}, \bm{Z}_{sj}, \sum_{i=0}^{m_s} Y_{sji}=1 }\left[ (1+ \sum \nolimits_{i=0}^{m_s-1} \exp \{ \beta_x(X_{sji}-X_{sjm_s}) + \bm{\beta}_z^T(\bm{Z}_{sji} -\bm{Z}_{sjm_s} )\})^{-1} \right],
\end{split}
\end{equation*}

\noindent where the surrogacy assumption is used in the third line. A second order Taylor series approximation about $\bm{X}_{sj}$ with respect to $E[\bm{X}_{sj}| \bm{W}_{sj}, \bm{Z}_{sj}, \sum_{i=0}^{m_s} Y_{sji}=1]$ yields the approximate contribution by stratum $j$ of study $s$, namely

\begin{equation*}
\begin{split}
\tilde{l}_{sj}=\left(1+ \sum_{i=0}^{m_s-1} \exp \{ \beta_x(\tilde{X}_{sji}-\tilde{X}_{sjm_s}) + \bm{\beta}_z^T(\bm{Z}_{sji} -\bm{Z}_{sjm_s})\} \right)^{-1},
\end{split}
\end{equation*}

\noindent where $\tilde{X}_{sji}-\tilde{X}_{sjm_s}=E[X_{sji}| W_{sji}, \bm{Z}_{sji}, \sum_{i=0}^{m_s} Y_{sji}=1]-E[X_{sjm_s}| W_{sjm_s}, \bm{Z}_{sjm_s}, \sum_{i=0}^{m_s} Y_{sji}=1]$.  This approximate conditional likelihood performs best when the elements in the variance-covariance matrix $Var(\bm{X}_{sj}|\bm{W}_{sj},\bm{Z}_{sj},\sum_{i=0}^{m_s} Y_{sji}=1)$ are small (i.e. there is minimal noise in the relationship between $X_{sji}$ and $W_{sji}$ conditional on the covariates and matching scheme) or when the association between $Y_{sji}$ and $X_{sji}$ is not strong. Further details about these conditions and the full derivation of the approximate conditional likelihood are available in Section 1 of the Supplementary Materials. The presence of $E[X_{sji}|W_{sji},\bm{Z}_{sji},\sum_{i=0}^{m_s} Y_{sji}=1]$ in the approximate conditional likelihood contribution $\tilde{l}_{sj}$ motivates the construction of calibration models.

\subsection{Calibration model}

We formulate study-specific calibration models among studies using a local laboratory to evaluate $\tilde{l}_{sj}$ above under the aggregated approaches. In each study using a local lab, we estimate a calibration model from a subset of controls who have both reference and local laboratory measurements available. The resulting model quantifies the relationship between $X_{sji}$ and $W_{sji}$ and is used to estimate the reference laboratory measurement among individuals with only a local laboratory measurement. 

For presentational simplicity, we drop $\bm{Z}_{sji}$ from the calibration models such that \\ $E[X_{sji}|W_{sji},\bm{Z}_{sji},\sum_{i=0}^{m_s} Y_{sji}=1] \approx E[X_{sji}|W_{sji},\sum_{i=0}^{m_s} Y_{sji}=1]$. Previous work demonstrates that, when not conditional on $\sum_{i=0}^{m_s} Y_{sji}=1$, omitting $\bm{Z}_{sji}$ has a negligibly small impact on the resulting calibration models so long as $\bm{Z}_{sji}$ is not very strongly correlated with $X_{sji}$ \citep{sloan2018design}. The additional condition, $\sum_{i=0}^{m_s} Y_{sji}=1$, does not change this conclusion.

We assume a linear relationship between the reference and local laboratory measurements among the matched cases and controls such that

\begin{equation} 
\label{calform}
E[X_{sji}|W_{sji}, \sum_{i=0}^{m_s} Y_{sji}=1]=a_s + b_s W_{sji}, 
\end{equation}

\noindent where $a_s$ and $b_s$ are the calibration parameters. Note that model (\ref{calform}) indicates that the calibration parameters are study-specific and invariant over the strata of each study. The assumption of invariance over strata could be relaxed by instead assuming $E[X_{sji}|W_{sji},\bm{U}_{sji},\sum_{i=0}^{m_s} Y_{sji}=1] = a_s + b_s W_{sji} + \bm{c}^T_{s} \bm{U}_{sji}$, where $\bm{U}_{sji}$ is a vector that may contain some or all of the matching factors. Previous work suggests that the simple linear model in (\ref{calform}) is appropriate and sufficient in most settings \citep{sloan2018design,gail2016calibration}.

Due to concerns about the availability of case biospecimens for re-assay at the reference laboratory, the calibration models in (\ref{calform}) are typically fit among controls only, i.e. $\hat{E}[X_{sji}|W_{sji},Y_{sji}=0]=\hat{a}_{s,co}+\hat{b}_{s,co} W_{sji}$. While $\hat{a}_{s,co}$ and $\hat{b}_{s,co}$ are generally not consistent estimators of  $a_s$ and $b_s$, we can show that the controls-only parameters approximate the parameters of model (\ref{calform}) under some relatively mild conditions. Specifically, under multivariate normality of $(\bm{X}_{sj},\bm{W}_{sj})$, $\hat{a}_{s,co}$ converges to a value close to $a_s$ when $E[\bm{X}_{sj}|\sum_{i=0}^{m_s} Y_{sji}=1] \approx E[\bm{X}_{sj}|\bm{Y}_{sj}=\bm{0}]$ and $Var(\bm{X}_{sj}|\sum_{i=0}^{m_s} Y_{sji}=1) \approx Var(\bm{X}_{sj}|\bm{Y}_{sj}=\bm{0})$, and $\hat{b}_{s,co}$ converges to a value close to $b_s$ when $Var(\bm{X}_{sj}|\sum_{i=0}^{m_s} Y_{sji}=1) \approx Var(\bm{X}_{sj}|\bm{Y}_{sj}=\bm{0})$. In addition, if the biomarker effect is small (that is, $\beta_x \approx 0$), then $\hat{a}_{s,co}$ and $\hat{b}_{s,co}$ also converge to values close to $a_s$ and $b_s$. Further mathematical details are available in Section 2 of the Supplementary Materials. We also illustrate the approximation for a single study via simulation, using simulation parameters listed in section 3.1. Over 1000 simulation replicates, the mean percent bias of the estimated parameters in (\ref{calform}) is close to zero while the controls-only calibration parameter estimates are only depressed up to twelve percent for very large RRs (Figure 1). The intercept estimates are typically more biased than the slope estimates, and the downward bias is more pronounced for increasingly non-null RRs. Simulations for other values of $a_s$ and $b_s$ (not shown here) demonstrate that these patterns are invariant to the specific values of the calibration parameters.

Let $\tilde{X}_{sji}$ be the estimated biomarker measurement used in place of $X_{sji}$ in the aggregated analysis. Its value depends on the chosen calibration approach and availability of the reference laboratory measurement. For studies using the reference laboratory for all biomarker measurements, no calibration occurs and $\tilde{X}_{sji}=X_{sji}$ for all study participants. For studies using a local lab, the \textit{full calibration} and \textit{internalized} methods are defined as follows:

\begin{equation*}
\begin{split}
\textrm{Full calibration method: } & \tilde{X}_{sji} = \hat{E}[X_{sji}|W_{sji}, Y_{sji}=0] \\
\textrm{Internalized method: } & \tilde{X}_{sji} = \begin{cases}
X_{sji}, & \textrm{if reference laboratory measurement available}\\
\hat{E}[X_{sji}|W_{sji}, Y_{sji}=0], & \textrm{otherwise,} \\
\end{cases}
\end{split}
\end{equation*}

\noindent where $\hat{E}[X_{sji}|W_{sji}, Y_{sji}=0]=\hat{a}_{s,co}+\hat{b}_{s,co} W_{sji}$. The pooled analysis uniformly applies one of these methods to all studies using local laboratories.

The biomarker measurements appear in the likelihood via the difference $X_{sji}-X_{sjm}$. For the full calibration method, the calibrated difference simplifies to $\tilde{X}_{sji}-\tilde{X}_{sjm}=\hat{b}_{s,co}(W_{sji}-W_{sjm})$ for all observations. Consequently, bias in the intercept $\hat{a}_{s,co}$ does not impact the $\beta_x$ estimator under the full calibration method. In contrast, for observations in the calibration study subset under the internalized method, $\tilde{X}_{sji}-\tilde{X}_{sjm}= X_{sji} - (\hat{a}_{s,co}+\hat{b}_{s,co}W_{sji})$. Thus, bias in both the intercept and slope estimates $\hat{a}_{s,co}$ and  $\hat{b}_{s,co}$ induces bias in the internalized method $\beta_x$ estimate. We will see in simulation studies that the relative performance of the internalized, full calibration, and two-stage methods is impacted by the size of the calibration subset.

\subsection{Parameter estimation}

Define $\bm{\beta}=[\beta_x, \bm{\beta}_z]$, $\bm{a}=[a_1,\dotsc , a_Q]$, and $\bm{b}=[b_1 , \dotsc ,b_Q]$ such that our target of estimation is $\bm{\theta}=[\bm{a} , \bm{b} , \bm{\beta}]$. We define estimating equations $ [\bm{\psi}_{\bm{a}} , \bm{\psi}_{\bm{b}} , \psi_{\beta_x} , \bm{\psi}_{\bm{\beta}_z} ] = \bm{0}$, where $\bm{\psi}_{\bm{a}}$, $\bm{\psi}_{\bm{b}}$,  $\psi_{\beta_x}$, and $\bm{\psi}_{\bm{\beta}_z}$ are approximately unbiased estimating functions for their respective parameters (see Section 3 of the Supplementary Materials). 

Estimates of $\bm{\beta}$ are obtained through a two step process involving the pseudo maximum likelihood estimation method \citep{gong1981pseudo}. We first estimate the calibration parameters $\hat{\bm{a}}$ and $\hat{\bm{b}}$ by fitting the linear regression calibration models. We then obtain $\hat{\bm{\beta}}$ using pseudo-maximum conditional likelihood estimation over the set of estimating equations $[\psi_{\beta_x}(\hat{\bm{a}},\hat{\bm{b}}) , \bm{\psi}_{\bm{\beta}_z}(\hat{\bm{a}},\hat{\bm{b}}) )]=\bm{0}$, where instances of $\bm{a}$ and $\bm{b}$ have been replaced with their estimates. Since the estimating equations for $\bm{\beta}$ are with respect to $\tilde{l}_{sj}$ where $\hat{\bm{a}}$ and $\hat{\bm{b}}$ are treated as known, the estimates are not the same as those obtained from maximizing a joint likelihood, which is the product of the likelihood for $\bm{\beta}$ and the likelihood for $\bm{a}$ and $\bm{b}$. 

We must account for uncertainty in the estimation of $\hat{\bm{a}}$ and $\hat{\bm{b}}$ when estimating $\widehat{Var}(\hat{\bm{\beta}})$. As a result, we use the sandwich variance formulas over the entire set of estimating equations. The appropriate diagonal terms of the sandwich variance matrix provide variance estimates for $\hat{\bm{\beta}}$ (see Section 4 of the Supplementary Materials).

\subsection{Two-stage approach}

The two-stage approach for pooled data uses regression calibration, a broadly applicable method initially developed in the measurement error literature, to adjust for calibration in the first stage study-specific analyses \citep{Carroll:1012043,rosner1990correction,spiegelman1997regression,spiegelman2001efficient}. The second stage combines these estimates using fixed effects meta-analysis.

Regression calibration for cohort studies using standard logistic regression has been robustly developed by Rosner, Spiegelman, and Willett, among others \citep{rosner1989correction, rosner1990correction, Carroll:1012043}. However, there has been less focus on the application of regression calibration to the setting of matched data with conditional logistic regression. Using simulation studies and randomly sampled internal validation studies, \cite{mcshane2001covariate} tested the performance of the regression calibration point estimate formula to obtain $\beta_x$ point estimates under conditional logistic regression. Their results demonstrated that regression calibration performed comparably to other measurement error methods for exposure effects that were not too strong. Effectively, the matched nature of the data is ignored in the calibration step but accounted for in the regression step. Further simulations by \cite{guolo2008simulation} established the ready extension of regression calibration to the matched data setting. However, neither manuscript provided simulations specific to the setting of controls-only calibration data, which we consider analytically in Supplementary Materials Section 2 and via simulation in Section 3. 

The assumptions required by regression calibration for matched case-control studies are similar, including: (i) the surrogacy assumption defined in section 2.1, and (ii) the noise in the calibration model is moderate and/or the association between $Y$ and $X$ is not too strong. 
    
We briefly describe the two-stage method and its implementation of regression calibration. For studies using the reference lab for all biomarker measurements, we directly obtain effect and variance estimates from a conditional logistic regression model. Among studies whose biomarkers were measured by local lab, we estimate $\hat{\beta}_{x,q}$, the adjusted effect estimate in study $q$ obtained from the regression calibration method such that

\begin{equation}
\begin{split}
\hat{\beta}_{x,q} & = \hat{\beta}_{w,q}/ \hat{b}_q, \\
\widehat{Var}(\hat{\beta}_{x,q}) & = \hat{b}_q^{-2} \widehat{Var}(\hat{\beta}_{w,q}) + \hat{b}_q^{-4} \hat{\beta}_{w,q}^2 \widehat{Var}(\hat{b}_q),  \\
\end{split}
\end{equation}

\noindent where $\hat{b}_q$ is obtained from the calibration study and $\hat{\beta}_{w,q}$ and $\widehat{Var}(\hat{\beta}_{w,q})$ are the naive estimates from a conditional logistic regression model fit using all local laboratory measurements of the individuals in study $q$. Formula (2.3) for $\widehat{Var}(\hat{\beta}_{x,q})$ leverages the fact that the conditional logistic regression estimates, $\hat{\beta}_{w,q}$, and the calibration parameter estimate, $\hat{b}_q$ are uncorrelated asymptotically (proof in Section 5 of Supplementary Materials). Fixed-effects meta-analysis then combines the study-specific estimates using inverse variance weights to obtain a final effect and variance estimate for $\hat{\beta}_x$.

\subsection{Two-stage method for models with an interaction term}

Suppose $V_{sji}$ is a covariate whose interaction with $X_{sji}$ is of interest. The conditional logistic regression model with an interaction term between $X_{sji}$ and $V_{sji}$ is

\begin{equation} 
\text{logit}(P(Y_{sji}=1|X_{sji}, V_{sji}, \bm{Z}_{sji}))= \beta_{0sj} + \beta_xX_{sji}+\beta_vV_{sji}+\beta_{xv}X_{sji}V_{sji}+\bm{\beta}_z^T \bm{Z}_{sji}.
\end{equation}

\noindent For the aggregated methods, the development of an approximate conditional likelihood involving an interaction term follows similarly to section 2.1 (see Section 1 of the Supplementary Materials). However, the two-stage method for models involving interaction terms must be modified to address inference for $(\beta_x, \beta_v, \beta_{xv})$. In $q^{th}$ study using a local laboratory, suppose we fit the following naive model using all local laboratory biomarker measurements:

\begin{equation*}
\begin{split}
\textrm{logit}(P(Y_{qji}=1|W_{qji},V_{qji}, \bm{Z}_{qji}))=\beta_{0qj}^{*} + \beta_{w,q}^{*}W_{qji} + \beta_{v,q}^{*}V_{qji} + \beta_{wv,q}^{*}W_{qji}V_{qji} + \bm{\beta}^{T*}_{z,q} \bm{Z}_{qji}.
\end{split}
\end{equation*}

\noindent Like before, assume we obtain study-specific estimates $\hat{a}_q$ and $\hat{b}_q$ by fitting the linear model $E[X_{qji}|W_{qji},Y_{qji}=0]=\hat{a}_q + \hat{b}_q W_{qji}$ in each calibration subset. We derive a modified form of regression calibration to obtain adjusted point estimates of $\beta_x$, $\beta_v$, and $\beta_{xv}$ in each study using a local laboratory such that

\begin{equation*}
\hat{\beta}_{x,q}   = \frac{\hat{\beta}_{w,q}^{*}}{\hat{b}_q}, \hspace{10mm} \hat{\beta}_{v,q}   = \hat{\beta}_{v,q}^{*}- \frac{\hat{a}_q \hat{\beta}_{wv,q}^{*}}{\hat{b}_q}, \hspace{10mm} \hat{\beta}_{xv,q}   = \frac{\hat{\beta}_{wv,q}^{*}}{\hat{b}_q}. \\
\end{equation*}

\noindent The full derivation of the point and variance estimates can be found in Supplement Section 6. Of note, the formulas for the point and variance estimates of $\hat{\beta}_{x,q}$ follow similarly to the case of no interaction term. The estimates for $\hat{\beta}_{xv,q}$ follow similarly to $\hat{\beta}_{x,q}$, and $\hat{\beta}_{v,q}$ has a slightly more complex adjustment given the interaction term between $X_{sji}$ and $V_{sji}$.

\section{Simulations}
\label{sec3}

\subsection{Model without an interaction term}

We simulated 1:1 matched case-control data by generating data for each stratum and then selecting a case and control at random from each stratum. Let $e_{sji}$ be the error term in the linear regression of $X_{sji}$ on $W_{sji}$. We assumed joint multivariate normality of  $(X_{sji}, W_{sji}, e_{sji})$ and generated data such that

\begin{equation*}
\begin{split}
\begin{pmatrix} X_{sji} \\ W_{sji} \\ e_{sji} \end{pmatrix} & \sim \textrm{MVN} \left( \begin{pmatrix} \mu_x \\ (\mu_x-a_s)/b_s \\ 0 \end{pmatrix},   \begin{pmatrix} \sigma_x^2 & b_s\sigma_{ws}^2 & \sigma_x^2-b_s^2\sigma_{ws}^2 \\ \cdot & \sigma_{ws}^2  & 0 \\ \cdot & \cdot & \sigma_x^2-b_s^2\sigma_{ws}^2 \end{pmatrix} \right).  \\
\end{split}
\end{equation*}

\noindent This formulation induces $E[X_{sji}|W_{sji}]=a_s+b_sW_{si}$ and $Cov(W_{sji}, e_{sji})=0$. We supposed that the stratum-specific intercepts $\beta_{0sj}$ were normally distributed such that $\beta_{0sj} \sim N(\mu_{\beta_0},\sigma^2_{\beta_0})$, and used a simple risk model without covariates, $\textrm{logit}(P(Y_{sji}=1|X_{sji})) = \beta_{0sj} + \beta_x X_{sji}$. 

We set $\mu_x=0$, $\sigma_x^2=1$,  $\mu_{\beta_0}=-1$ and $\sigma^2_{\beta_0}$=0.01. We assumed four studies contributed to the final analysis, each with 1000 total subjects (or 500 case-control pairs) and 100 calibration participants, and assigned calibration parameters $\bm{a}=[-3,1,-1,3]$, and $\bm{b}=[0.5,0.75,1.25,1.5]$. We completed 1000 simulation replicates at six RRs, including $(1.25, 1.5, 1.75, 2, 2.25, 2.5)$.

We compared the two aggregated procedures and the two-stage approach with regard to mean percent bias over the simulation replicates, mean squared error (MSE), and coverage rate, defined as the proportion of simulation replicates whose 95\% confidence intervals covered the true $\beta_x$. For comparison, we also included a naive method that does not calibrate and uses $W_{sji}$ in the logistic regression. We assumed that increasing levels of the biomarker increased disease risk, although similar results were obtained when the biomarker was instead protective against disease. We provide tables of results in the main paper and include graphs of the mean percent bias estimates from these tables in Section 7 of the Supplementary Materials.

As shown in Table 1, the naive method performed poorly over all effects and underscored the need for calibration. At every RR considered, the percent bias of the naive estimate exceeded -27\% and the corresponding coverage rates were less than 40\%. The full calibration method estimates were consistently less biased than the internalized method estimates. Over the range of RRs considered, the internalized estimate was biased downward by approximately three to four percent while the full calibration estimate was biased by less than one percent. For large RRs greater than or equal to 2.0, the coverage rate under the internalized method was 90\% to 91\% due to depression of point estimates. In contrast, the coverage rates of the full calibration and two-stage methods ranged from 93\% to 96\%. The two-stage approach generally performed comparably to the full calibration method with regards to percent bias in the RR estimate, MSE, and coverage rate.

We also performed simulations that fixed the total sample size at 1000 participants while varying the calibration subset size between 30, 50, and 150 subjects (or 3\%, 5\%, and 15\% participation rates, respectively). As shown in Figure 2, at all calibration study sizes, the full calibration method offered nearly unbiased point estimates. However, the internalized method estimates experienced increasing downward bias as the proportion of subjects participating in the calibration subset increased owing to increasingly differential calibration of cases and controls. As calibration study size increased, the two-stage method point estimates were increasingly less biased owing to the improved estimation of calibration parameters.

\subsection{Model with an interaction term}

For the simulations involving a model with an interaction term, we also generated $V_{sji}$ in the multivariate normal model for each stratum such that

\begin{equation*}
\begin{split}
\begin{pmatrix} W_{sji} \\ V_{sji} \\ e_{sji} \end{pmatrix} & \sim \textrm{MVN} \left( \begin{pmatrix} \mu_{ws} \\ \mu_v \\ 0 \end{pmatrix},   \begin{pmatrix} \sigma^2_{ws} &  \sigma_{wvs} & 0 \\ \cdot & \sigma_v^2 & 0   \\ \cdot & \cdot & \sigma^2_e \end{pmatrix} \right),  \\
X_{sji} & = a_s +b_sW_{sji} +e_{sji},
\end{split}
\end{equation*}

\noindent where $\mu_{ws} = (\mu_x-a_s)/b_s$, $\sigma_{wvs} = Cov(W_{sji},V_{sji})$, and $\sigma_{ws}^2 =(\sigma_x^2-\sigma^2_e)/b_s^2$. We again used 1:1 matching and omitted other covariates such that the risk model was $\text{logit}(P(Y_{si}=1|X_{sji},V_{sji})) = \beta_{0sj} + \beta_x X_{sji} + \beta_v V_{sji} + \beta_{xv} X_{sji}V_{sji}$. Like before, we assumed four studies contributed to the analysis, each with 1000 total subjects and 100 individuals in the calibration subset. We set $\mu_x=\mu_v=0$, $\sigma^2_x=\sigma_v^2=1$,  $\bm{a}=(-3,1,-1,3)$, $\bm{b}=(0.5, 0.75, 1.25, 1.5)$, and $\text{Corr}(X_{sji},V_{sji})=0.2$, which in turn induced $\sigma_{wvs}$ for each study. We considered the same range of RRs for the main effect of the biomarker measurement and also chose four combinations of $(\beta_{v},\beta_{xv})$ to address all possible qualitative effects of the covariate and interaction term, including $(\exp(\beta_v),\exp(\beta_{xv})) \in [(0.8,0.8),(0.8,1.2),(1.2,0.8),(1.2,1.2)]$. The simulation results for $(\exp(\beta_v),\exp(\beta_{xv}))=(1.2,1.2)$ are reported in Table 2 while the results for $(\exp(\beta_v),\exp(\beta_{xv})) \in [(0.8,1.2),(1.2,0.8),(1.2,1.2)]$ are presented in Section 7 of the Supplementary Material.

We focus first on the operating characteristics of $\hat{\beta}_x$ (Table 2, and Table 1 and Figure 2 of Section 7 of the Supplementary Materials). The mean $\hat{\beta}_x$ estimate under the internalized method is consistently biased downward by two to four percent across all effect sizes and all combinations of $(\exp(\beta_v),\exp(\beta_{xv}))$. The full calibration and two-stage point estimates were separated by no more than one percent and were approximately unbiased for the true RR. All methods performed markedly better than the naive approach, whose point biases ranged from -32 to -23 percent. The MSE of the two-stage and aggregated methods were similar across all effect combinations.

For $\beta_v$, the full calibration and two-stage approach also performed similarly with regard to percent bias (Table 2, and Table 2 and Figure 2 of Section 7 of the Supplementary Materials). The internalized method estimate was the least biased near the null of $\beta_x$ but became more biased for increasingly protective or deleterious effects. For example, when $(\beta_x, \beta_v, \beta_{xv})=(1.2,1.2,1.2)$, the percent bias of $\beta_v$ was 1.5\%, and when $(\beta_x, \beta_v, \beta_{xv})=(2.5,1.2,1.2)$, the percent bias of $\beta_v$ was 7.6\%. The naive estimates of $\beta_v$ were increasingly biased as $|\beta_x|$ increased, with biases reaching -28.3 and 23.0 percent in some settings. 

Across all combinations of effects, the percent bias in estimates of $\beta_{xv}$ from all three calibration methods were contained within (-4.9\%, 3.9\%) (Table 2, and Table 3 and Figure 2 of Section 7 of Supplementary Materials). The $\hat{\beta}_{xv}$ estimate under the internalized method was frequently more biased than the corresponding estimates under the full calibration or two-stage approaches, which performed similarly to each other. The naive method performed poorly and the percent bias in its $\beta_{xv}$ estimates ranged from -105.4\% to -81.7\%.

\section{Applied example}
\label{sec4}

We completed two data examples to illustrate the methods. In the first example, we investigate the impact of circulating 25-hydroxyvitamin D (25(OH)D) levels on risk of stroke. In the second example, we investigate the impact of 25(OH)D levels and its interaction with a dichotomized BMI term on the risk of a composite outcome, fatal or nonfatal stroke or myocardial infarction (henceforth referred to as the CVD endpoint). In both examples, we match each case to a single control based on sex and age at blood draw.

We applied the aggregated, two-stage, and naive methods to combine data derived from three large prospective cohort studies in the United States, including the HPFS \citep{wu2011interactions}, the NHS1 \citep{eliassen2016plasma}, and the NHS2 \citep{eliassen2011plasma}. The HPFS began enrollment in 1986 and includes 51,529 male health professionals aged 40 to 75 years at baseline. The NHS1 enrolled 121,701 female nurses aged 30 to 55 years at baseline in 1976. The NHS2, a younger counterpart to the NHS1, was established in 1989 with the enrollment of 116,671 female nurses, aged 25 to 42 years at baseline. In each cohort, participants completed biannual questionnaires providing information about medical history, diet, and lifestyle conditions; further general discussion of each cohort is available elsewhere \citep{grobbee1990coffee,colditz1997nurses,rich2002physical}. Between 1989 and 1997, each study completed laboratory assays on blood samples for a host of biomarkers, including 25(OH)D, from a subset of participants \citep{wu2011interactions, eliassen2016plasma,eliassen2011plasma}. Subjects with a previous cancer diagnosis were not eligible for random selection. Individuals were excluded from the pooled analysis if they did not have 25(OH)D measurements available or stroke or myocardial infarction outcome data.

Each study obtained calibration data among a subset of controls by re-assaying their blood samples at Heartland Assays, LLC (Ames, IA) during 2011-2013. Circulating 25(OH)D levels were standardized to 20 nmol/L increments. Information about the main studies and the calibration subsets, including the parameter estimates of the study-specific calibration models, is presented in Table 3.

In the first example involving the stroke endpoint, we pooled 179 matched case-control pairs. Previous work by \cite{sun201225} showed that individuals with 25(OH)D measurements in the top tertile of the population had reduced risk of stroke compared to individuals with measurements in the bottom tertile. Our analyses coarsely matched on age (grouped into tertiles in each cohort) and adjusted for years of follow-up after blood draw, smoking status (never/ever), family history of myocardial infarction (yes/no), personal history of hypertension (yes/no), BMI (less than or greater or equal to 25 kg/m$^2$), and personal history of diabetes (yes/no). The internalized, full calibration, and two-stage methods all demonstrate that increased 25(OH)D levels correlate with a small, non-significant protective effect against stroke, with RRs of 0.95, 0.96, and 0.95 respectively (Table 6). The naive approach, which pools all local laboratory measurements directly without adjustment, yielded a RR of 0.99 and a confidence interval that was narrower than those estimated under the calibration methods.

In the second example involving the CVD endpoint, we pooled 624 case-control pairs. Some literature suggests that vitamin D deficiency is more deleterious among individuals with high BMI \citep{weng2013vitamin,levi2017vitamin}. We dichotomized BMI into a binary variable with value 1 if the subject is not overweight (BMI$<$25 kg/m$^2$) and value 0 otherwise. In addition to the matching factors, analyses were adjusted for years of follow-up after blood draw, smoking status (never/ever), family history of myocardial infarction (yes/no), personal history of hypertension (yes/no), and personal history of diabetes (yes/no). The estimated regression coefficients and confidence intervals for 25(OH)D, BMI, and their interaction are presented in Table 5. All pooling methods indicated that having higher circulating 25(OH)D and a non-overweight BMI were associated with a lower risk of stroke. Based on the results in Table 5, the RR from the aggregated and two-stage methods associated with a 20 nmol/L increase in 25(OH)D among subjects with a BMI less than 25 kg/m$^2$ ranged from 0.85 to 0.86, while the RR from the naive analysis was 0.78. Among overweight subjects, the corresponding RR from the calibration methods ranged from 0.91 to 0.92, and the naive RR estimate was 0.84 (see Table 4 of Section 7 in the Supplementary Materials). As this illustrative example demonstrates, the bias in the naive RR estimate for models including interaction term is not necessarily towards the null. The bias may be toward the alternative hypotheses and thus lead to false-positive scientific findings.

\section{Discussion}
\label{sec5}

In this work, we proposed statistical methods for calibrated biomarker data pooled across multiple nested case-control studies. Our methods facilitate inference on the main effect of the biomarker as well as a biomarker-covariate interaction term. Keeping with common practice, we estimated study-specific calibration models from subsets of controls re-assayed at the reference lab.

Several recommendations and observations stem from our work. Regardless of whether one is interested in the main effect of the biomarker or the interaction term, the full calibration approach is the preferred aggregated method for calibrating biomarker measurements that were measured using different laboratories and/or assays. Average percent bias in $\beta_x$ and $\beta_{xv}$ estimates are minimized by the full calibration method. Due to using a controls-only calibration scheme, we observed that $(\hat{a}_{s,co},\hat{b}_{s,co})$ were slightly biased for the parameters in model (\ref{calform}) for strong biomarker effects. Any bias in these estimates is uniformly incorporated in both cases and controls under the full calibration approach such that bias is minimized in the $\beta_x$ and $\beta_{xv}$ estimates. In fact, the intercept $\hat{a}_{s,co}$ cancels out in the approximate conditional likelihood contribution for the full calibration approach, which does not occur for the internalized method. 

Under the two-stage method, we also derived estimators with analytic variances for conditional logistic regression models including an interaction term between the biomarker exposure and a potential effect modifier. We consistently observed that the full calibration and two-stage methods offered similar point estimates, standard errors, and coverage rates. In simulation, the difference in effect estimates between the full calibration and two-stage methods was less than two percent regardless of the inclusion of an interaction term. When incorporating an interaction term, the direction of bias in $\beta_v$ was not consistent and depended on the direction and magnitude of $\beta_x$ and $\beta_v$. Naive estimates were typically quite biased and illustrated the risk of failing to implement a calibration step when necessary. More problematically, the naive estimates were sometimes biased toward the alternative, resulting in an inflated type I error rate.

Although the aggregated and two-stage methods are equally viable and valid options for analyzing pooled data, logistical considerations may dictate the preferred approach for the statistical analysis. For instance, aggregated methods often lend themselves better to subgroup analyses because they reduce issues resulting from data sparsity. If the main exposure effect and at least some covariate effects are homogeneous, the aggregated method may also offer efficiency gains in covariate estimation relative to the two-stage method \citep{lin2010relative}. However, the two-stage method may be more appealing than the aggregated methods at times for its intuitive and simple implementation, and its robustness to these covariate homogeneity assumptions.

\section{Software}
\label{sec6}

Functions in the form of R code is available at the first author's Github account and \\ \url{https://www.hsph.harvard.edu/molin-wang/software}.

\section{Supplementary Material}
\label{sec7}

Supplementary material is available online at \url{http://biostatistics.oxfordjournals.org}. Section 1 derives the approximate likelihood; Section 2 discusses the calibration parameters under controls-only calibration; Sections 3 and 4 give the estimating equations and sandwich variance estimator respectively; Section 5 demonstrates zero covariance between the naive estimate from conditional logistic regression and calibration parameter estimates; Section 6 derives formulas for the two-stage method with an interaction term; Section 7 gives additional figures and tables.

\section*{Acknowledgments}

We are grateful to Tao Hou and Shiaw-Shyuan (Sherry) Yaun for their assistance in accessing the data. We also thank the Circulating Biomarkers and Breast and Colorectal Cancer Consortium team (R01CA152071, PI: Stephanie Smith-Warner; Intramural Research Program, Division of Cancer Epidemiology and Genetics, National Cancer Institute: Regina Ziegler) for conducting the calibration study in the vitamin D examples.  A.S. was supported by NIH grant T32-NS048005 and M. W. was supported by NIH/NCI grant R03CA212799. {\it Conflict of Interest}: None declared.

\bibliographystyle{biorefs}
\bibliography{refs}

\newpage

\begin{figure}[h]
\centering
\includegraphics[width=12cm]{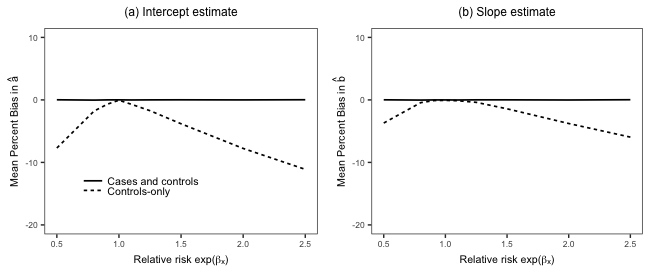}
\caption{Comparison of percent bias in average calibration parameter estimates over 1000 simulation replicates for a single study's calibration parameters when fitting the model $E[X_{sji}|W_{sji}, \sum_{i=0}^{m_s} Y_{sji}=1]=\hat{a}_s + \hat{b}_s W_{sji}$, where $(a_s, b_s)=(3, 0.8)$ in the data generating mechanism. The model parameters in (2.2) result from fitting a calibration model in the cases and controls of the nested case-control study, while the controls-only parameters result from fitting a calibration model among only controls.}
\end{figure}

\begin{figure}[h]
\centering
\includegraphics[width=12cm]{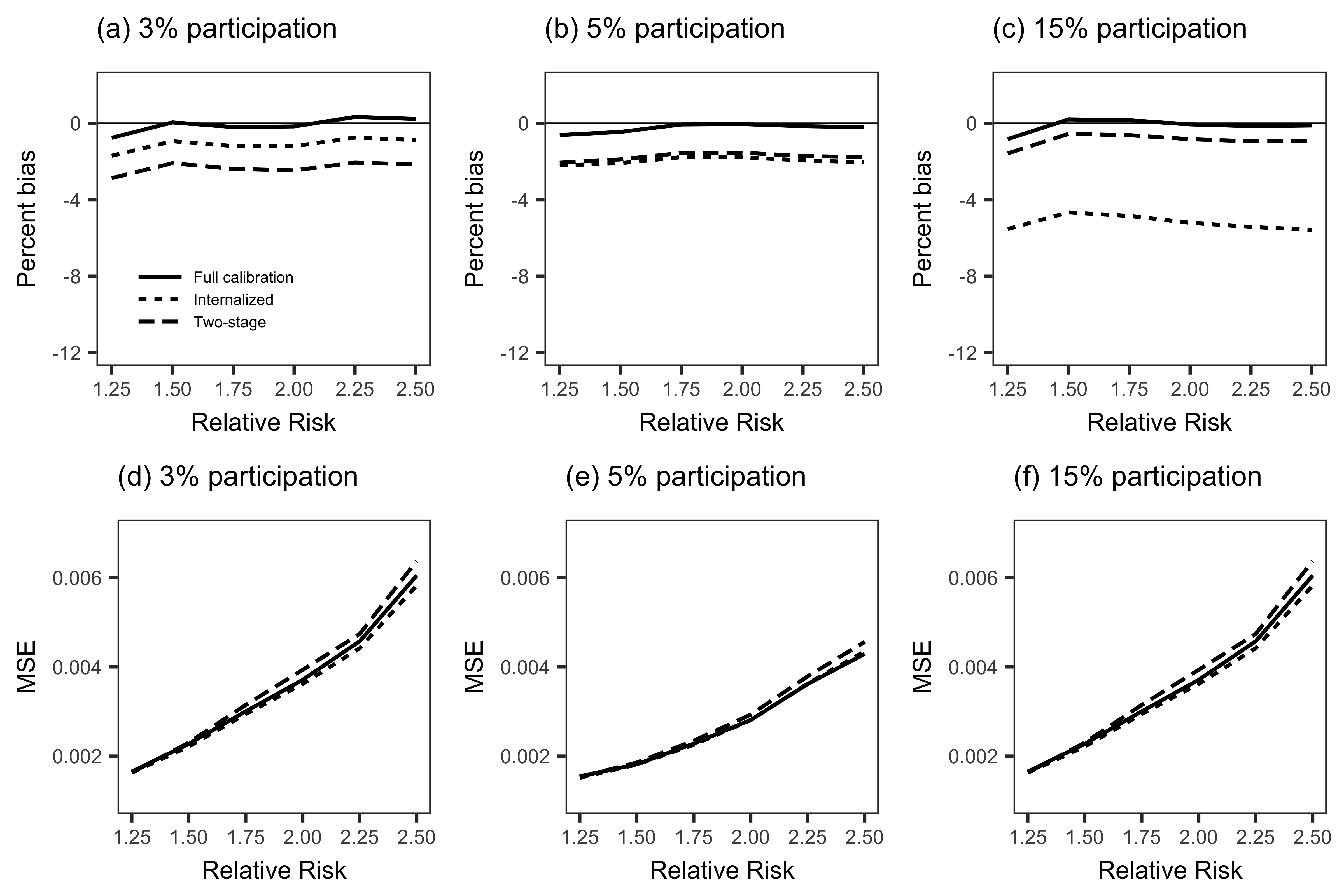}
\caption{Comparison of methods as number of participants in the calibration study increases. The number of subjects in each study remains fixed at 1000, or equivalently, 500 case-control pairs. The calibration study participation rates considered are 3\%, 5\%, and 15\%, or 30, 50, and 150 individuals respectively.}
\end{figure}

\newpage

\begin{sidewaystable}[p]
\centering
\caption{\textit{Comparison of operating characteristics for $\beta_x$ under the model $\textrm{logit}(P(Y_{sji}=1|X_{sji})) = \beta_{0sj} + \beta_xX_{sji}$, for naive ($\hat{\beta}_{N}$), internalized ($\hat{\beta}_{IN}$), full calibration ($\hat{\beta}_{FC}$), and two-stage ($\hat{\beta}_{TS}$) methods. Percent bias and MSE are computed by $(\hat{\beta}-\beta)/\beta$ and $(\beta - 
\hat{\beta})^2$ respectively and the reported value is the average over 1000 simulations. Standard error is the square root of the empirical variance over all replicates. The coverage rate is the proportion of simulations whose estimated 95\% confidence intervals covered the true effect $\beta_x$.}
\label{Table1}}
{\tabcolsep=4pt
\begin{tabular}{lcccclcccclcccc}
\hline
                          & \multicolumn{4}{c}{Mean percent bias (SE)}                 &  & \multicolumn{4}{c}{MSE}           &  & \multicolumn{4}{c}{Coverage rate} \\ \cline{2-5} \cline{7-10} \cline{12-15} 
\multicolumn{1}{c}{$\beta_x$} & $\hat{\beta}_N$ & $\hat{\beta}_{IN}$ & $\hat{\beta}_{FC}$ & $\hat{\beta}_{TS}$ &  & $\hat{\beta}_N$ & $\hat{\beta}_{IN}$ & $\hat{\beta}_{FC}$ & $\hat{\beta}_{TS}$ &  & $\hat{\beta}_N$ & $\hat{\beta}_{IN}$ & $\hat{\beta}_{FC}$ & $\hat{\beta}_{TS}$      \\ \hline
log(1.25)                 & -29.4 (0.029) & -3.1 (0.037) & 0.1 (0.038)  & -0.8 (0.038) &  & 0.0051 & 0.0014 & 0.0014 & 0.0014 &  & 0.37   & 0.96   & 0.95   & 0.96   \\
log(1.50)                  & -29.0 (0.032) & -3.2 (0.040) & 0.0 (0.042)  & -0.9 (0.042) &  & 0.0149 & 0.0018 & 0.0018 & 0.0018 &  & 0.05   & 0.95   & 0.94   & 0.95   \\
log(1.75)                 & -28.6 (0.035) & -3.5 (0.043) & -0.1 (0.045) & -1.0 (0.045) &  & 0.0269 & 0.0023 & 0.0021 & 0.0021 &  & 0.01   & 0.93   & 0.94   & 0.95   \\
log(2.00)                    & -28.2 (0.039) & -3.5 (0.049) & 0.0 (0.051)  & -1.0 (0.051) &  & 0.0396 & 0.0030 & 0.0027 & 0.0027 &  & 0.00   & 0.91   & 0.93   & 0.93   \\
log(2.25)                 & -28.0 (0.042) & -3.6 (0.052) & -0.1 (0.055) & -1.1 (0.055) &  & 0.0532 & 0.0036 & 0.0030 & 0.0031 &  & 0.00   & 0.90   & 0.94   & 0.94   \\
log(2.50)                  & -27.9 (0.044) & -3.9 (0.055) & -0.3 (0.058) & -1.3 (0.058) &  & 0.0671 & 0.0043 & 0.0034 & 0.0035 &  & 0.00   & 0.90   & 0.96   & 0.94   \\ \hline
\end{tabular}}
\end{sidewaystable}

\newpage

\newgeometry{top=35mm, bottom=30mm}
\begin{sidewaystable}[p]
\centering
 \caption{ \textit{  Comparison of operating characteristics for ($\beta_x$,$\beta_v$,$\beta_{xv}$) for naive (N), internalized (IN), full calibration (FC), and two-stage (TS) methods under the model $\text{logit}(P(Y_{sji}=1|X_{sji}, V_{sji}, \bm{Z}_{sji}))= \beta_{0sj} + \beta_xX_{sji}+\beta_vV_{sji}+\beta_{xv}X_{sji}V_{sji}$, with $(\exp(\beta_v),\exp(\beta_{xv}))=(1.2,1.2)$.}}
\begin{tabular}{ccccclcccclcccc}
\hline
              & \multicolumn{4}{c}{Percent bias (SE) of $\beta_x$}         &  & \multicolumn{4}{c}{MSE of $\beta_x$} &  & \multicolumn{4}{c}{Coverage rate of $\beta_x$} \\ \cline{2-5} \cline{7-10} \cline{12-15} 
$\beta_x$ & N             & IN           & FC           & TS           &  & N       & IN      & FC      & TS     &  & N          & IN        & FC        & TS        \\ \hline
$\log(0.5)$           & -25.8 (0.036) & -2.9 (0.045) & 0.9 (0.048)  & 0.4 (0.048)  &  & 0.0342  & 0.0023  & 0.0023  & 0.0023 &  & 0.00       & 0.95      & 0.96      & 0.96      \\
$\log(0.8)$           & -29.4 (0.031) & -2.7 (0.037) & 0.6 (0.039) & 0.2 (0.039) &  & 0.0050  & 0.0015  & 0.0015  & 0.0015 &  & 0.41       & 0.96      & 0.96      & 0.95      \\
$\log(1.2)$           & -28.6 (0.030) & -2.8 (0.037) & 0.3 (0.038)  & -0.1 (0.038)  &  & 0.0038  & 0.0014  & 0.0015  & 0.0014 &  & 0.54       & 0.96      & 0.96      & 0.96      \\
$\log(1.5)$           & -27.6 (0.032) & -2.8 (0.041) & 0.4 (0.043)  & -0.2 (0.043) &  & 0.0140  & 0.0018  & 0.0018  & 0.0018 &  & 0.06       & 0.96      & 0.95      & 0.95      \\
$\log(2.0)$          & -25.7 (0.040) & -2.6 (0.048) & 0.9 (0.050)  & 0.8 (0.050)  &  & 0.0333  & 0.0026  & 0.0026  & 0.0026 &  & 0.00       & 0.95      & 0.96      & 0.95      \\
$\log(2.5)$           & -23.4 (0.045) & -2.5 (0.054) & 1.4 (0.057)  & 1.1 (0.057)  &  & 0.0478  & 0.0034  & 0.0035  & 0.0034 &  & 0.00       & 0.96      & 0.96      & 0.95      \\ \hline
              & \multicolumn{4}{c}{Percent bias (SE) of $\beta_v$}         &  & \multicolumn{4}{c}{MSE of $\beta_v$} &  & \multicolumn{4}{c}{Coverage rate of $\beta_v$} \\ \cline{2-5} \cline{7-10} \cline{12-15} 
$\beta_x$ & N             & IN           & FC           & TS           &  & N       & IN      & FC      & TS     &  & N          & IN        & FC        & TS        \\ \hline
$\log(0.5)$           & -26.9 (0.035)   & -6.5 (0.034)  & -1.0 (0.035) & -1.0 (0.035) &  & 0.0019  & 0.0012  & 0.0012  & 0.0012 &  & 0.88       & 0.96      & 0.95      & 0.95      \\
$\log(0.8)$           & -15.6 (0.034)  & -2.7 (0.033)  & -0.9 (0.034)  & -1.0 (0.034)  &  & 0.0014  & 0.0011  & 0.0011  & 0.0011 &  & 0.93       & 0.95      & 0.95      & 0.94      \\
$\log(1.2)$           & -0.8 (0.033) & 1.5 (0.033)  & 0.3 (0.033)  & 0.3 (0.033)  &  & 0.0011  & 0.0011  & 0.0011  & 0.0011 &  & 0.95       & 0.95      & 0.95      & 0.95      \\
$\log(1.5)$           & 5.9 (0.034) & 3.1 (0.033) & 0.3 (0.033) & 0.2 (0.033) &  & 0.0012  & 0.0011  & 0.0011  & 0.0011 &  & 0.95       & 0.96      & 0.96      & 0.96      \\
$\log(2.0)$           & 12.1 (0.035) & 5.4 (0.035) & 0.6 (0.035)  & 0.4 (0.035)  &  & 0.0013  & 0.0012  & 0.0012  & 0.0012 &  & 0.94       & 0.96      & 0.96      & 0.96      \\
$\log(2.5)$           & 15.0 (0.035) & 7.6 (0.037) & 1.3 (0.037)  & 1.2 (0.037)  &  & 0.0015  & 0.0014  & 0.0014  & 0.0014 &  & 0.94       & 0.95      & 0.95      & 0.95      \\ \hline
              & \multicolumn{4}{c}{Percent bias (SE) of $\beta_{xv}$}         &  & \multicolumn{4}{c}{MSE of $\beta_{xv}$} &  & \multicolumn{4}{c}{Coverage rate of $\beta_{xv}$} \\ \cline{2-5} \cline{7-10} \cline{12-15} 
$\beta_x$ & N             & IN           & FC           & TS           &  & N       & IN      & FC      & TS     &  & N          & IN        & FC        & TS        \\ \hline
$\log(0.5)$           & -82.8 (0.011) & 1.3 (0.042) & 2.3 (0.043)  & 1.6 (0.043)  &  & 0.0063  & 0.0018  & 0.0019  & 0.0018 &  & 0.00       & 0.95      & 0.95      & 0.96      \\
$\log(0.8)$           & -88.7 (0.010) & 0.4 (0.038) & 0.9 (0.038) & 0.3 (0.038) &  & 0.0072  & 0.0014  & 0.0015  & 0.0015 &  & 0.00       & 0.95      & 0.95      & 0.95      \\
$\log(1.2)$           & -93.9 (0.010) & 0.4 (0.037)  & 0.0 (0.038)  & -0.7 (0.038) &  & 0.0081  & 0.0014  & 0.0014  & 0.0014 &  & 0.00       & 0.95      & 0.95      & 0.95      \\
$\log(1.5)$           & -96.6 (0.010) & 1.9 (0.039)  & 0.3 (0.039)  & -0.4 (0.039) &  & 0.0086  & 0.0015  & 0.0015  & 0.0015 &  & 0.00       & 0.95      & 0.95      & 0.95      \\
$\log(2.0)$           & -99.0 (0.011) & 0.7 (0.044)  & -2.1 (0.044) & -2.7 (0.044) &  & 0.0090  & 0.0019  & 0.0020  & 0.0020 &  & 0.00       & 0.95      & 0.96      & 0.95      \\
$\log(2.5)$           & -100.3 (0.013) & 3.2 (0.048) & -1.1 (0.049) & -1.9 (0.048) &  & 0.0093  & 0.0023  & 0.0024  & 0.0023 &  & 0.00       & 0.95      & 0.96      & 0.95      \\ \hline
\end{tabular}
\end{sidewaystable}
\restoregeometry

\newpage

\begin{table}[h]
\centering
\caption{\textit{Number of case-control pairs $(N)$, size of the calibration study $(n_{cal})$, and the estimated intercept $(\hat{a})$ and slope $(\hat{b})$ of the calibration model for each study participating in the pooled analysis.}}
\begin{tabular}{ccccclcccc}
\hline
      & \multicolumn{4}{c}{Model 1: Stroke endpoint}        &  & \multicolumn{4}{c}{Model 2: CVD endpoint}           \\ \cline{2-5} \cline{7-10} 
Study & $N$ & $n_{cal}$ & $\hat{a}$ (SE) & $\hat{b}$ (SE) &  & $N$ & $n_{cal}$ & $\hat{a}$ (SE) & $\hat{b}$ (SE) \\ \hline
HPFS  & 49    & 25        & 4.35 (2.80)    & 0.94 (0.04)    &  & 406   & 18        & 5.33 (3.72)    & 0.93 (0.05)    \\
NHS1  & 103   & 27        & 6.50 (2.88)    & 0.84 (0.04)    &  & 175   & 27        & 6.50 (2.88)    & 0.84 (0.04)    \\
NHS2  & 27    & 28        & 9.01 (4.88)    & 0.95 (0.07)    &  & 43    & 28        & 9.01 (4.88)    & 0.95 (0.07)    \\
Total & 179   & 80        &                &                &  & 624   & 73        &                &                \\ \hline
\end{tabular}
\end{table}

\begin{table}[h!]
\centering
\caption{\textit{Point and RR estimates for the association of circulating 25(OH)D* and stroke, adjusting for years of follow-up after blood draw, BMI (overweight or not), smoking (never/ever), family history of myocardial infarction (yes/no), hypertension (yes/no), and diabetes (yes/no).}}
\begin{tabular}{cccc}
\hline
          Method       & $\hat{\beta}_x$  & RR    & RR 95\% CI        \\ \hline
Internalized     & -0.051 & 0.950 & (0.721, 1.253) \\ 
Full calibration & -0.046 & 0.955 & (0.715, 1.276) \\
Two-stage     & -0.048 & 0.953 & (0.717, 1.266) \\ 
Naive & -0.018 & 0.983 & (0.833, 1.159) \\ \hline
\end{tabular}
\\
\textit{*Estimates correspond to a 20 nmol/L increase in circulating 25(OH)D}
\end{table}

\begin{table}[h!]
\centering
\caption{\textit{Point estimates with 95\% confidence intervals for circulating 25(OH)D,  BMI*, and their interaction with CVD as the outcome event, adjusting for years of follow-up after blood draw, smoking (never/ever), family history of myocardial infarction (yes/no), hypertension (yes/no), and diabetes (yes/no). }}
\begin{tabular}{cccc}
\hline
         Method        & 25(OH)D                 & BMI                & BMI $\times$ 25(OH)D              \\ \hline
Internalized     & -0.091 (-0.232, 0.050) & -0.592 (-0.663, -0.521) & -0.072 (-0.267, 0.123)  \\
Full calibration & -0.089 (-0.246, 0.068) & -0.587 (-0.661, -0.513) & -0.069 (-0.271, 0.133) \\
Two-stage        & -0.086 (-0.244, 0.072) & -0.569 (-0.643, -0.495) & -0.066 (-0.266, 0.135) \\
Naive            & -0.173 (-0.380, 0.032) & -0.163 (-0.374, 0.053) & -0.082 (-0.180, 0.016) \\ \hline
\end{tabular}
\\
\textit{*BMI is treated as a dichotomized variable taking value 1 if less than $25 kg/m^2$ and 0 otherwise}
\end{table}

\end{document}